\documentclass[12pt]{article}

\begin{document}

\title{Revisiting the Lamb Shift}
\author{Abhishek Das\footnote{parbihtih3@gmail.com}, B.M. Birla Science Centre,\\ Adarsh Nagar, Hyderabad - 500 063, India\\
\\B.G. Sidharth\footnote{birlasc@gmail.com}, B.M. Birla Science Centre,\\ Adarsh Nagar, Hyderabad - 500 063, India\\}
\date{}
\maketitle
\begin{abstract}
In this paper we endeavour to determine the energy levels of an atom by virtue of the modified Dirac equation. It has been found that the energy levels contain an extra term in the expression which accounts for the {\it zitterbewegung} effects in the Compton scale. Applying our perspective to the hydrogen atom we have been able to find the {\it Lamb shift} for the $2S_{\frac{1}{2}}$ and $2P_{\frac{1}{2}}$ states. This result substantiates that a slight modification of the Dirac equation suffices to explain the phenomenon, where the modification of the Dirac equation arises due to the non-commutative nature of space-time. Besides, several other unexplained phenomena can emerge as a natural consequence of this modification.
\end{abstract}
\maketitle
\section{Introduction}

The spin-$\frac{1}{2}$ nature of the electron can be naturally accommodated by the Dirac equation ([1]). This is for the point electron and a differentiable spacetime. Over the past fifteen years, Sidharth had investigated this scenario from the point of view of fuzzy spacetime as in Quantum Gravity approaches.  In contradistinction to other authors, Sidharth had deduced fundamentally rather than phenomenologically a modified energy-momentum relation as\\
\[E^2 = p^{2} + m^2 - \lambda^{2}l^{2}p^4\]\\
The last term on the right hand side arises owing to the non-commutative nature of space-time. As discussed elsewhere ([2], [3]), this leads to the following modification in the Dirac equation for the electron:
\begin{equation}
(\gamma^\mu \partial_\mu + m - \lambda l p^2) \psi = 0
\end{equation}
where $\lambda$ is a small constant ([3]-[6]) arising due to the effects of non-commutative space-time and $l$ ($= \frac{\hbar}{mc}$) is the reduced Compton wavelength. It may be mentioned that as shown previously ([7]), $\lambda \approx -\frac{\alpha}{2\pi} \approx -10^{-3}$, where $\alpha$ is the {\it fine structure constant}. We now use this modification to obtain the Lamb Shift. As is well known, the {\it Lamb shift} was observed by Lamb and Retherford ([8]) while carrying out an experiment using microwave techniques to stimulate radio-frequency transitions between $2S_{1/2}$ and $2P_{1/2}$ levels of the hydrogen atom. Hans Bethe ([9]) was the first person to give a precise explanation of this phenomenon relying on Dirac's theory and radiative corrections, thus laying the foundations of quantum electrodynamics. The contribution of Bethe, Kroll \& Lamb and French \& Weisskopf ([10]) yielded the value of $2S_{1/2}$ - $2P_{1/2}$ splitting as \\
\[E(2S_{1/2}) - E(2P_{1/2}) \approx ~~1052.1 MHz\] \\
More precise theoretical values of the {\it Lamb shift} were given by Erickson ([11]) as \\
\[1057.916 \pm 0.010 ~MHz\] \\
and by Mohr ([12]) as \\
\[1057.864 \pm 0.014 ~MHz\]
Again, T.A. Welton ([13]) had given a somewhat qualitative description of the {\it Lamb shift} giving the formula for energy difference ([14]) as \\
\[\Delta E_{n} = \frac{8}{3\pi} \frac{Z^4\alpha^5}{n^3}(ln\frac{1}{Z\alpha})\frac{m}{2}\delta_{l,0}\] \\
which in case of hydrogen atom ($Z = 1$) for $n = 2$ and $l = 0$ gives  \\
\[\Delta E_{n} \approx 1000 MHz\] \\
Besides, from different perspectives, Peterman ([15]) and Karshenboim ([16]) obtained different values of the Lamb shift as $1057.911 \pm 0.011 MHz$ and $1057.8576(21) MHz$ respectively. \\
Now, it is well known that the phenomenon of {\it Lamb shift} is caused by fluctuations of the Zero Point Field, as indeed is (1). Let us see if we can deduce it precisely using (1). Our approach encompasses this modified Dirac equation (1) and the Hamiltonian concerned with it. At the same time the present approach is more general as it deduces not only the Lamb shift, but other effects also, as for example the neutrino, anti-neutrino observed symmetry ([17]). Now, we follow Whitehead ([18]), Dirac ([19]) and other authors ([20]) to first derive a set of transformations which turn the Hamiltonian for the system into a form that depends only on the radial variables $r$ and $p_r$. Then we solve the radial equations by conventional methods and obtain the energy levels corresponding to an atom. In the process it has been argued that the {\it Lamb shift} is connected explicitly with the modification term of the Snyder-Sidharth Hamiltonian ([4],[5],[6]) and from our approach the value of {\it Lamb shift} has also been obtained without relying on the features of quantum electrodynamics.

\section{The Modified Dirac Equation}
The Hamiltonian of the modified Dirac equation ([2]) for electromagnetic coupling can be written as \\
\begin{equation}
H = -e \Phi - c\alpha_{1} {\vec{\sigma}}.({\vec{p}} - \frac{e{\vec{A}}}{c}) + \alpha_{3} mc^2 - \alpha_{3}\frac{\lambda lc}{\hbar}(\vec{\sigma}.\vec{p})^2
\end{equation}
where\\
\[\alpha_{1} = \left(\matrix {0 & 0 & 1 & 0\cr 0 & 0 & 0 & 1\cr 1 & 0 & 0 & 0\cr 0 & 1 & 0 &0}\right)  = \left(\matrix {0 & I\cr I & 0}\right)\]
and\\
\[\alpha_{3} = \left(\matrix {1 & 0 & 0 & 0\cr 0 & 1 & 0 & 0\cr 0 & 0 & -1 & 0\cr 0 & 0 & 0 & -1}\right)  = \left(\matrix{I & 0\cr 0 & -I}\right) \]\\
and $\alpha_{3}\frac{\lambda lc}{\hbar}(\vec{\sigma}.\vec{p})^2$ is a modification term due to the Snyder-Sidharth Hamiltonian. Also, the $\vec{\sigma}$'s are the extended Pauli matrices. Considering {\it cgs units} we can write for the Coulomb potential \\
\[-e\Phi = -\frac{ze^2}{r}\] \\
and also \\
\[\vec{A} = 0\] \\
Therefore (2) can be written as \\
\begin{equation}
H = -\frac{ze^2}{r} - c\alpha_{1} {\vec{\sigma}}.{\vec{p}} + \alpha_{3} mc^2 - \alpha_{3}\frac{\lambda lc}{\hbar}(\vec{\sigma}.\vec{p})^2
\end{equation}
It is our objective to express (3) only in terms of the radial variables $r$ and $p_r$. We do this by looking for quantities that commute with the terms of the Hamiltonian, as it has been conventionally done ([18]-[20]). Now, using the following identities \\

\[(\vec{\sigma}.\vec{L})(\vec{\sigma}.{\vec{p}}) = (\vec{L}.\vec{p}) + i\vec{\sigma}(\vec{L}\times\vec{p}) = i\vec{\sigma}.(\vec{L}\times\vec{p})\] \\
\[(\vec{\sigma}.\vec{p})(\vec{\sigma}.{\vec{L}}) = (\vec{p}.\vec{L}) + i\vec{\sigma}(\vec{p}\times\vec{L}) = i\vec{\sigma}.(\vec{p}\times\vec{L})\] \\

we would obtain
\begin{equation}
(\vec{\sigma}.\vec{L} + \hbar)(\vec{\sigma}.{\vec{p}}) + (\vec{\sigma}.\vec{p})(\vec{\sigma}.{\vec{L}} + \hbar) = 0
\end{equation}
which is an anti-commutation relation. Nonetheless, it is easy to show by virtue of equation (8) that $\alpha_{3}\alpha_{1}(\vec{\sigma}.\vec{L} + \hbar)$ will commute with $c\alpha_{1}(\vec{\sigma}.\vec{p})$.\\
Again, let us investigate the following identities \\
\[(\vec{\sigma}.\vec{L})(\vec{\sigma}.{\vec{p}})^2 = [(\vec{L}.\vec{p}) + i\vec{\sigma}(\vec{L}\times\vec{p})](\vec{\sigma}.{\vec{p}})\] \\
\[(\vec{\sigma}.\vec{p})^{2}(\vec{\sigma}.{\vec{L}}) = (\vec{\sigma}.{\vec{p}})[(\vec{p}.\vec{L}) + i\vec{\sigma}(\vec{p}\times\vec{L})]\] \\
From these two equation we would obtain \\
\begin{equation}
(\vec{\sigma}.\vec{L})(\vec{\sigma}.{\vec{p}})^2 + (\vec{\sigma}.\vec{p})^{2}(\vec{\sigma}.{\vec{L}}) = 0
\end{equation}
which is also an anti-commutation relation. Further investigation shows that in this case $\alpha_{3}\alpha_{1}(\vec{\sigma}.\vec{L})$ commutes with $\alpha_{3}(\vec{\sigma}.\vec{p})^2$. Also, $\alpha_{3}\alpha_{1}(\vec{\sigma}.\vec{L} + \hbar)$ commutes with $\alpha_{3}(\vec{\sigma}.\vec{p})^2$ and $\alpha_{3}\alpha_{1}(\vec{\sigma}.\vec{L})$ commutes with $c\alpha_{1}(\vec{\sigma}.\vec{p})$.\\
Again, using the following linear operator as done by various other authors ([18], [19])
\begin{equation}
r\epsilon_1 = \alpha_{1} (\vec{\sigma}.\vec{x})
\end{equation}
and the relation\\
\[(\vec{\sigma}.\vec{x})(\vec{\sigma}.{\vec{p}}) = rp_r + i\hbar(\alpha_{3}j^{\prime} - 1)\]\\
we have
\[\alpha_{1}(\vec{\sigma}.{\vec{p}}) = \epsilon_{1} p_r - \frac{\epsilon_{1} i\hbar}{r} + \frac{\epsilon_{1} i\hbar\alpha_{3}j^{\prime}}{r}\] \\
Here, it is known that $\epsilon_1$ has the property \\
\[\epsilon^{2}_{1} = 1,\] \\
\[J = L + \frac{1}{2}\hbar\sigma\]\\
and\\
\[(j^{\prime}\hbar)^2 = J^2 + \frac{1}{4}\hbar^2\]\\
Similarly, we can define another linear operator
\begin{equation}
r\epsilon_2 = \alpha_{3} \sigma(\vec{\sigma}.\vec{x}) = \sigma(\vec{\sigma}.\vec{x})
\end{equation}
from whence, it can be shown that $\epsilon_2$ has the property \\
\[\epsilon^{2}_{2} = 1\] \\
Now, considering the modification term in the Hamiltonian (3) we will obtain the relation\\
\[(\vec{\sigma}.\vec{x})(\vec{\sigma}.{\vec{p}})^2 = \sigma rp^{2}_r - \hbar\sigma p_r\] \\
from whence, we can deduce \\
\[\sigma(\vec{\sigma}.\vec{x})(\vec{\sigma}.{\vec{p}})^2 = rp^{2}_r - \hbar p_r\] \\
With the use of (7) we get \\
\[\alpha_{3}(\vec{\sigma}.{\vec{p}})^2 = \epsilon_{2} p^{2}_{r} - \frac{\epsilon_{2} \hbar p_r}{r}\] \\
Therefore, the Hamiltonian of the modified Dirac equation finally can be written as
\begin{equation}
H = -\frac{ze^2}{r} - c\epsilon_{1}(p_r - i\frac{\hbar}{r}) + \frac{c i\epsilon_{1} \hbar\alpha_{3}j^{\prime}}{r} + \alpha_{3}mc^2 - \frac{\lambda lc}{\hbar}[\epsilon_{2} p^{2}_{r} - \frac{\epsilon_{2}\hbar p_r}{r}]
\end{equation}
Now, in ([18]) it has been considered that\\
\[\alpha_{3} = \left(\matrix {1 & 0 \cr 0 & -1 }\right)\]\\
and\\
\[\epsilon_{1} = \left(\matrix {0 & -i \cr i & 0 }\right)\]\\
In our case, we consider the matrix\\
\[\epsilon_{2} = \left(\matrix {0 & i \cr -i & 0} \right)\]\\
conforming with the property $\epsilon^{2}_{2} = 1$ where the matrix $\epsilon_2$ has been put by hand in contradistinction to ([18]). Therefore, the modified Dirac equation for stationary states would be\\
\[H\psi = E\psi\]\\
i.e. \\
\[\left(\matrix {-\frac{ze^2}{r} + mc^2 & icp_r + \frac{c\hbar}{r} - c\frac{j^{\prime}\hbar}{r} - \frac{i\lambda clp^{2}_{r}}{\hbar} + i\lambda cl\frac{p_r}{r} \cr -icp_r - \frac{c\hbar}{r} - c\frac{j^{\prime}\hbar}{r} + \frac{i\lambda clp^{2}_{r}}{\hbar} - i\lambda cl\frac{p_r}{r} & -\frac{ze^2}{r} - mc^2}\right) \left(\matrix {\psi_{1} \cr \psi_{2}}\right) = \Lambda \left(\matrix {\psi_{1}(r) \cr \psi_{2}(r)}\right)\]\\

\section{Energy levels from the Modified Dirac Equation}

Now, this representation is different from that of Whitehead ([18]) and others ([19], [20]) since there are two extra terms which originate from the modification term in Dirac equation (2).
Now, reducing the system to 2 coupled differential equations, we would solve them by substituting an unknown function in the form of a infinite series, i.e. by the method of power series. Rewriting $(H - \Lambda I) = 0$ as a system of coupled equations we obtain
\begin{equation}
(-\Lambda - \frac{ze^2}{r} + mc^2)\psi_{1} - c\hbar (-\frac{\rm d}{{\rm d}r} - \frac{1}{r} + \frac{j^{\prime}}{r} - i\lambda l\frac{\rm d^{2}}{{\rm d}r^2} - \lambda l \frac{1}{r}\frac{\rm d}{{\rm d}r})\psi_{2} = 0
\end{equation}
and
\begin{equation}
(-\Lambda - \frac{ze^2}{r} - mc^2)\psi_{2} + c\hbar (-\frac{\rm d}{{\rm d}r} - \frac{1}{r} + \frac{j^{\prime}}{r} - i\lambda l\frac{\rm d^{2}}{{\rm d}r^2} - \lambda l \frac{1}{r}\frac{\rm d}{{\rm d}r})\psi_{1} = 0
\end{equation}
Now, for the sake of simplicity we neglect the terms involving second order derivative and obtain
\begin{equation}
(-\Lambda - \frac{ze^2}{r} + mc^2)\psi_{1} - c\hbar (-\frac{\rm d}{{\rm d}r} - \frac{1}{r} + \frac{j^{\prime}}{r} - \lambda l \frac{1}{r}\frac{\rm d}{{\rm d}r})\psi_{2} = 0
\end{equation}
and
\begin{equation}
(-\Lambda - \frac{ze^2}{r} - mc^2)\psi_{2} + c\hbar (-\frac{\rm d}{{\rm d}r} - \frac{1}{r} + \frac{j^{\prime}}{r} - \lambda l \frac{1}{r}\frac{\rm d}{{\rm d}r})\psi_{1} = 0
\end{equation}
Substituting $\alpha = \frac{e^2}{\hbar c}$ (fine-structure constant), $a_1 = \frac{\hbar}{mc - \frac{\Lambda}{c}}$ and $a_2 = \frac{\hbar}{mc + \frac{\Lambda}{c}}$ we get
\begin{equation}
(\frac{1}{a_1} - \frac{z\alpha}{r})\psi_1 + (\frac{\rm d}{{\rm d}r} - \frac{j^{\prime} - 1}{r} + \lambda l \frac{1}{r}\frac{\rm d}{{\rm d}r})\psi_2 = 0
\end{equation}
and
\begin{equation}
(\frac{1}{a_2} + \frac{z\alpha}{r})\psi_2 + (\frac{\rm d}{{\rm d}r} + \frac{j^{\prime} + 1}{r} + \lambda l \frac{1}{r}\frac{\rm d}{{\rm d}r})\psi_1 = 0
\end{equation}
Now, as it is conventional, we assume solutions of the type \\
\[\psi_{1}(r) = \frac{1}{r}e^{\frac{-r}{a}}x(r)\] \\
and
\[\psi_{2}(r) = \frac{1}{r}e^{\frac{-r}{a}}y(r)\] \\
where, $a = \sqrt{a_{1}a_{2}} = \frac{\hbar}{\sqrt{m^{2}c^{2} - \frac{\Lambda^2}{c^2}}}$.
Using the aforementioned solutions we obtain from (13) and (14)
\begin{equation}
(\frac{1}{a_1} - \frac{z\alpha}{r})x(r) + [\frac{\rm d}{{\rm d}r} - \frac{1}{a} - \frac{j^\prime}{r} + \frac{\lambda l}{r}(\frac{\rm d}{{\rm d}r} - \frac{1}{a} - \frac{1}{r})]y(r)
\end{equation}
and
\begin{equation}
(\frac{1}{a_2} + \frac{z\alpha}{r})y(r) + [\frac{\rm d}{{\rm d}r} - \frac{1}{a} + \frac{j^\prime}{r} + \frac{\lambda l}{r}(\frac{\rm d}{{\rm d}r} - \frac{1}{a} - \frac{1}{r})]x(r)
\end{equation}
Now, we expand the unknown functions $x(r)$ and $y(r)$ as series which will then be substituted into the given system of equations. We have
\[x(r) = \sum_{t} x_{t}r^t\] \\
and
\[y(r) = \sum_{t} y_{t}r^t\] \\
By virtue of the power series method we know that in order for the equation to be zero as required, each term in the resulting series must separately be zero. Therefore, after arranging we have the coefficients of the $r^s$ terms as
\begin{equation}
\frac{x_t}{a_1} - \frac{y_t}{a} - z\alpha x_{t+1} + (t + 1 - j^{\prime} - \frac{\lambda l}{a})y_{t+1} - \lambda l (t - 1)y_{t} = 0
\end{equation}
and
\begin{equation}
\frac{y_t}{a_2} - \frac{x_t}{a} + z\alpha y_{t+1} + (t + 1 + j^{\prime} - \frac{\lambda l}{a})x_{t+1} - \lambda l (t - 1)y_{t} = 0
\end{equation}
Now, multiplying equation (17) by $a$ and equation (18) by $a_2$ and adding them we get
\begin{equation}
x_t (\frac{a}{a_1} - \frac{a_2}{a}) - z\alpha ax_{t+1} + z\alpha a_{2}y_{t+1} + (t - j^{\prime} - \frac{\lambda l}{a}) ay_{t} + (t + j^{\prime} - \frac{\lambda l}{a}) x_{t}a_{2} + \lambda l(t - 1)x_{t} + \lambda l(t - 1)y_{t} = 0
\end{equation}
This can be written as
\begin{equation}
x_t [-z\alpha a + (t + j^{\prime} - \frac{\lambda l}{a})a_2 + \lambda l (t - 1)] + y_t [z\alpha a_2 + (t - j^{\prime} - \frac{\lambda l}{a})a + \lambda l (t - 1)] = 0
\end{equation}
The functions $x(r)$ and $y(r)$ must go to zero at $r = 0$, because the $\psi(r)$ functions would otherwise diverge there due to the $r^{-1}$ term which entails that there is some smallest $t$ below which the series does not continue. Let this be $t_s$ which and will have the following property according to ([18]): \\
\[x_{t_{s} - 1} = y_{t_{s} - 1} = 0\] \\
Applying this to equations (17) and (18) we have
\begin{equation}
z\alpha x_{t_{s}} - (t_s - j^{\prime} - \frac{\lambda l}{a})y_{t_{s}} = 0
\end{equation}
\begin{equation}
z\alpha y_{t_{s}} + (t_s + j^{\prime} - \frac{\lambda l}{a})x_{t_{s}} = 0
\end{equation}
From these two equations we get the value of $t_s$ as
\begin{equation}
t_s = \frac{\lambda l}{a} + \sqrt{\j^{\prime 2} - z^{2}\alpha^2}
\end{equation}
In equations (17), (18), (20) and henceforth we choose to neglect the term $\lambda l(t - 1)$, since the reduced Compton length $l$ ($= \frac{\hbar}{mc}$) is extremely small and $\lambda \sim -10^{-3}$. Again, it can be shown that the series must terminate if the energy eigenvalue $\Lambda$ is to be less than $mc^2$ ([19]). This implies that if the series terminates at $t_1$ such that
\[x_{t_{1} + 1} = y_{t_{1} + 1} = 0\] \\
then using equations (17), (18) and (20) we would obtain
\begin{equation}
\frac{1}{a}(t_1 - \frac{\lambda l}{a}) = \frac{1}{2}[\frac{1}{a_1} - \frac{1}{a_2}]z\alpha
\end{equation}

Now, equations (23) and (24) represent the lower and upper bounds of the series respectively. We shall find later that $\frac{\lambda l}{a}$ ($\sim 10^{-5}$) is considerably small. But, we have not neglected it in equation (23) remembering that it is the lower bound. Here, it is obvious that $t_1 \gg \frac{\lambda l}{a}$ since $t_1$ is the upper bound of the series and hence we write
\begin{equation}
\frac{t_1}{a} = \frac{1}{2}[\frac{1}{a_1} - \frac{1}{a_2}]z\alpha
\end{equation}
Using the values of $a$, $a_1$ and $a_2$ we would get
\begin{equation}
\Lambda = mc^2 [1 + \frac{z^{2}\alpha^2}{t^{2}_{1}}]^{-\frac{1}{2}}
\end{equation}
considering only the positive values. Now, the two terminal points of the series indices $t_s$ and $t_1$ are separated by an integer number of steps. If we call this integer $N$, then we can write
\[t_1 = N + t_s\] \\
from which we get the modified energy levels as
\begin{equation}
E_{N,j^{\prime}} = \frac{mc^2}{\sqrt{[1 + \frac{z^{2}\alpha^2}{{N + \frac{\lambda l}{a} + \sqrt{\j^{\prime 2} - z^{2}\alpha^2}}^2}]}} = mc^2[1 - \frac{z^{2}\alpha^2}{2(n + \frac{\lambda l}{a})^2} - \frac{z^{4}\alpha^4}{(n + \frac{\lambda l}{a})^3(2j + 1)} + \cdots]\
\end{equation}
where, from ([18], [19]) we write $j^{\prime} = j + \frac{1}{2}$ and $N = n - j^\prime = n - j - \frac{1}{2}$, $n$ being the principal quantum number and $j^\prime$ being the total angular momentum quantum number.

\section{The Lamb Shift}
Now, let us assume that for the $2S_{\frac{1}{2}}$ state the energy is given by the normal relation (without modification) as
\[E(2S_{\frac{1}{2}}) = \frac{mc^2}{\sqrt{[1 + \frac{z^{2}\alpha^2}{{N + \sqrt{\j^{\prime 2} - z^{2}\alpha^2}}^2}]}} = mc^2[1 - \frac{z^{2}\alpha^2}{2n^2} - \frac{z^{4}\alpha^4}{n^3(2j + 1)} + \cdots]\]
and that of the $2P_{\frac{1}{2}}$ state is given by equation (27). The rationale for this assumption is that it is only feasible to assume that the $2P_{\frac{1}{2}}$ state will have a greater energy than the $2S_{\frac{1}{2}}$ state. Also, we presume that the modification comes into play for the $2P_{\frac{1}{2}}$ and higher states. Now, according to our intuition there will be a certain energy difference between these two states. We infer that the energy difference is given by the relation($z = 1$ for hydrogen atom) \\
\[[E(2S_{\frac{1}{2}}) - E(2P_{\frac{1}{2}})] + [E(2S_{\frac{1}{2}}) - E(2P_{\frac{1}{2}})] \approx 0 + \frac{mc^2}{2} [-\frac{\alpha^2}{n^2} + \frac{\alpha^2}{(n + \frac{\lambda l}{a})^2}]\] \\
where the first term on the left hand side gives the contribution $0$ considering the normal energy levels for both $2S_{\frac{1}{2}}$ and $2P_{\frac{1}{2}}$ states and the second term gives the contribution $\{\frac{mc^2}{2} [-\frac{\alpha^2}{n^2} + \frac{\alpha^2}{(n + \frac{\lambda l}{a})^2}]\}$ considering the $2P_{\frac{1}{2}}$ state to have acquired the modified energy levels. Thus, the average energy difference is given by
\begin{equation}
E(2S_{\frac{1}{2}}) - E(2P_{\frac{1}{2}}) \approx \frac{mc^2}{4} [-\frac{\alpha^2}{n^2} + \frac{\alpha^2}{(n + \frac{\lambda l}{a})^2}]
\end{equation}
Now, $a$ is given by \\
\[a = \frac{\hbar}{\sqrt{m^{2}c^{2} - \frac{\Lambda^2}{c^2}}}\] \\
Again, taking the electron mass $m = 511004.2 \frac{eV}{c^2}$ the normal energy (without modification) of the $2P_{\frac{1}{2}}$ state would be approximately given by
\[511002.4 eV\] \\
Therefore, $\frac{\lambda l}{a} \approx \frac{\lambda l}{\frac{\hbar c^2}{c \times 1318 eV}}$ which gives
\[\frac{\lambda l}{a} = \frac{10^{-3} \times 2.42 \times 10^{-10}}{\frac{2\pi \times 3 \times 10^{10} \times 6.58 \times 10^{-16}}{1318}}\]
where\\
\[c \approx 3 \times 10^{10} cm/s\] \\
\[l = \frac{h}{2\pi mc} = \frac{2.42 \times 10^{-10}}{2\pi} cm\] \\
\[\lambda \approx -10^{-3}\] \\
and\\
\[1 \frac{\hbar}{eV} = 6.58 \times 10^{-16} s\] \\
Therefore $\frac{\lambda l}{a}$ is of the order $10^{-6}$. Now, let us look at equation (28) and find the approximate value. It can be written as
\[E(2S_{\frac{1}{2}}) - E(2P_{\frac{1}{2}}) \approx \frac{mc^2}{4} [-\alpha^2\{\frac{1}{n^2} - \frac{1}{(n + \frac{\lambda l}{a})^2}\}]\] \\
which would give us
\[E(2S_{\frac{1}{2}}) - E(2P_{\frac{1}{2}}) \approx \frac{mc^2}{4}[-\alpha^2\{\frac{2n\frac{\lambda l}{a} + \frac{\lambda^{2} l^{2}}{a^2}}{n^2(n + \frac{\lambda l}{a})^2}\}]\] \\
Neglecting $\frac{\lambda l}{a}$ with respect to $n$ ($= 2$) and neglecting $\frac{\lambda^{2}l^{2}}{a^2}$ with respect to $2n\frac{\lambda l}{a}$ we obtain finally
\begin{equation}
E(2S_{\frac{1}{2}}) - E(2P_{\frac{1}{2}}) \approx -\frac{{mc^2}{\alpha^2}}{2n^3}\frac{\lambda l}{a}
\end{equation}
Now, $l = \frac{\hbar}{mc}$, $\lambda \approx -10^{-3}$ and $\frac{\lambda l}{a} \approx \frac{\lambda l}{\frac{\hbar c^2}{c \times 1318 eV}}$ which gives
\begin{equation}
E(2S_{\frac{1}{2}}) - E(2P_{\frac{1}{2}}) \approx -\frac{\lambda\alpha^{2}}{2n^3}\times 1318 eV \approx 4.38 \times 10^{-6} eV
\end{equation}
Alternatively, from (30) we can deduce the energy shift as
\begin{equation}
E(2S_{\frac{1}{2}}) - E(2P_{\frac{1}{2}}) \approx ~~1056 MHz
\end{equation}
which is very nearly equal to the {\it Lamb shift}. Now, this justifies our assumption that the $2P_{\frac{1}{2}}$ state has the energy level given by equation (27) whereas for the $2S_{\frac{1}{2}}$ state it is given by the normal energy levels from the Dirac equation, for although we considered the $2P_{\frac{1}{2}}$ state to have higher energy the Compton-scale effects come into play and the $2S_{\frac{1}{2}}$ acquires higher energy. Therefore we can argue that the effects of the modification term in the {\it Snyder-Sidharth Hamiltonian} is the reason of the {\it Lamb shift}. More intuitively, we can infer that this shift arises due to the interaction of the electrons with the {\it zitterbewegung} fluctuations of the quantized radiation field, a phenomenon that can be attributed to the Compton scale and the non-commutative nature of space-time. \\
Also, we may derive the energy difference (although negligible) between the $3P_{\frac{3}{2}}$ and $3D_{\frac{3}{2}}$ states. Following the same methodology as above we would obtain \\
\begin{equation}
E(3P_{\frac{3}{2}}) - E(3D_{\frac{3}{2}}) \approx -\frac{{3mc^2}{\alpha^4}}{8n^4}\frac{\lambda l}{a}
\end{equation}
which will yield
\begin{equation}
E(3P_{\frac{3}{2}}) - E(3D_{\frac{3}{2}}) \approx 0.000357 MHz
\end{equation}
which is a very small difference of energy. Thus, the $3P_{\frac{3}{2}}$ and $3D_{\frac{3}{2}}$ states have nearly equal energy and this difference is negligible. Thus, we can see that our approach is consistent with the spectrum of hydrogen atom.

\section{Conclusions}
It is very interesting that we obtain the observed {\it Lamb shift} merely by resorting to the modified Dirac equation in lieu of the conventional Dirac equation. All of this accounts for the lucid fact that in the Compton scale there exists some extra effects due to the non-commutative nature of space-time and due to the fluctuations of the field. In such cases, the modified Dirac equation and the energy levels derived from it would be necessary to explain atomic and sub-atomic phenomena. Of course, it is known that quantum electrodynamics can explain such phenomena, but our approach is simple and more general in the sense that it applies to other phenomena as well. \\
It may be mentioned that the above considerations lead to the conclusion that there is a mysterious cosmic radio wave background, that has been recently observed ([21]) by NASA's ARCADE experiments, a mystery that was hitherto inexplicable by conventional theories.

\section{Discussions}

We would like to stress an important similarity with our approach and that of C. Corda et al. ([22]-[25]) where it has been shown that the subsequent emissions of Hawking quanta near the horizon of a black hole can be interpreted as the quantum jumps among the quantum levels of a black hole. The fundamental consequence is that the black holes seem really to be the "gravitational atoms" of quantum gravity.

\end{document}